\documentclass[aps,prb,twocolumn,showpacs,groupedaddress]{revtex4}

\usepackage{graphicx}
\usepackage{latexsym}
\usepackage{amssymb}

\begin{document}

\title{Magnetism and superconductivity in $M_{c}$Ta$_{2}$S$_{2}$C ($M$ = Fe,
Co, Ni, and Cu)}

\author{Masatsugu Suzuki}
\email[]{suzuki@binghamton.edu}
\affiliation{Department of Physics, State University of New York at Binghamton, 
Binghamton, New York 13902-6000}

\author{Itsuko S. Suzuki}
\email[]{itsuko@binghamton.edu}
\affiliation{Department of Physics, State University of New York at Binghamton, 
Binghamton, New York 13902-6000}

\author{J\"{u}rgen Walter}
\email[]{juergen.walter@chemiemetall.de}
\altaffiliation{CM Chemiemetall GmbH, Niels-Bohr-Str. 5, Bitterfeld, 
Germany}
\affiliation{Department of Materials Science and Processing, Graduate
School of Engineering, Osaka University, 2-1, Yamada-oka, Suita, 565-0879,
JAPAN}


\date{\today}

\begin{abstract}
Magnetic properties of $M_{c}$Ta$_{2}$S$_{2}$C ($M$ = Fe, Co, Ni, Cu) have
been studied using SQUID DC and AC magnetic susceptibility.  In these
systems magnetic $M^{2+}$ ions are intercalated into van der Waals gaps
between adjacent S layers of host superconductor Ta$_{2}$S$_{2}$C.
Fe$_{0.33}$Ta$_{2}$S$_{2}$C is a quasi 2D $XY$-like antiferromagnet on the
triangular lattice.  It undergoes an antiferromagnetic phase transition at
$T_{N}$ (= 117 K).  The irreversible effect of magnetization occurs below
$T_{N}$, reflecting the frustrated nature of the system.  The AF phase
coexists with two superconducting phases with the transition temperatures
$T_{cu} = 8.8$ K and $T_{cl} = 4.6$ K. Co$_{0.33}$Ta$_{2}$S$_{2}$C is a
quasi 2D Ising-like antiferromagnet on the triangular lattice.  The
antiferromagnetic phase below $T_{N} = 18.6$ K coexists with a
superconducting phase below $T_{cu} = 9.1$ K. Both
Ni$_{0.25}$Ta$_{2}$S$_{2}$C and Cu$_{0.60}$Ta$_{2}$S$_{2}$C are
superconductors with $T_{cu}$ ($= 8.7$ K for Ni and 6.4 K for Cu) and
$T_{cl}$ (= 4.6 K common to $M_{c}$Ta$_{2}$S$_{2}$C).  Very small effective
magnetic moments suggest that Ni$^{2+}$ and Cu$^{2+}$ spins are partially
delocalized.
\end{abstract}


\maketitle



\section{\label{intro}INTRODUCTION}
$X_{2}$S$_{2}$C ($X$ = Ta and Nb) belongs to the class of layered
chalcogenides, where a sandwiched structure of $X$-C-$X$-S-S-$X$ is
periodically stacked along the $c$ axis.  The structure of $X_{2}$S$_{2}$C
can be viewed as a structural sum of $X$-C layer and $X$S$_{2}$ layered
structure.  There are very weak bonds between adjacent S layers, which is
called van der Waals (vdw) gap.  The structural and physical properties
have been studied by several research
groups.\cite{Beckmann1970,Wally1998a,Wally1998b,Walter2000a,Walter2000b,
Boller1992,Sakamaki2001,Walter2004,Wally1998c} These compounds show
superconductivity at low temperatures.  Various kinds of transition metals
$M$ (= Ti, V, Cr, Mn, Fe, Co, Ni, Cu) can be intercalated into the host
system $X_{2}$S$_{2}$C, forming intercalation compounds denoted by
$M_{c}X_{2}$S$_{2}$C, where $c$ is the concentration of $M$ atoms.  The
basic structure of the host crystal remains unchanged.  The structural and
magnetic properties of $M_{c}$Ta$_{2}$S$_{2}$C and $M_{c}$Nb$_{2}$S$_{2}$C
have been reported by Boller and Sobczak\cite{Boller1971} and Boller and
Hiebl,\cite{Boller1992} respectively.  The system $M_{c}X_{2}$S$_{2}$C with
$M$ = V, Cr, Mn, Fe, Co, and Ni has a $c$-axis stacking sequence denoted by
a 3R-$M_{c}X_{2}$S$_{2}$C with a periodicity of three $X$-C-$X$-S-$M$-S-$X$
sandwiched structures, where $M$ is located in octahedral sites in the vdw
gaps between adjacent S layers.  In contrast, Cu$_{c}X_{2}$S$_{2}$C has a
$c$-axis stacking sequence denoted by 1T-Cu$_{c}X_{2}$S$_{2}$C, where Cu
atoms are tetrahedrally coordinated by S atoms inside the S layers.  The
intercalate $M$ layer forms a regular triangular lattice (the lattice
constant $a$) in 3R-$M_{c}$Ta$_{2}$S$_{2}$C, which is either commensurate
or incommensurate with the in-plane structure of the host lattice.  Since
the in-plane structure of the host also forms a triangular lattice (the
lattice constant $a_{0} = 3.30 \AA$), the ratio of $a/a_{0}$ is related to
the concentration $c$ as $a/a_{0} = (1/c)^{1/2}$.  The nearest neighbor
distance $a$ is equal to $2a_{0}$ ($= 6.6 \AA$) for $c = 1/4$ forming a ($2
\times 2$) commensurate structure, and it is equal to $\sqrt{3}a_{0}$ ($=
5.72 \AA$) for $c = 1/3$ forming a ($\sqrt{3} \times \sqrt{3}$)
commensurate structure.

In our previous paper\cite{Walter2004} we have reported the
superconductivity of the pristine Ta$_{2}$S$_{2}$C. It undergoes successive
superconducting transitions of a hierarchical nature at $T_{cl} = 3.61 \pm
0.01$ K and $T_{cu} = 8.9 \pm 0.1$ K. The intermediate phase between
$T_{cu}$ and $T_{cl}$ is an intra-grain superconductive state occurring in
the TaC-type structure in Ta$_{2}$S$_{2}$C. The low temperature phase below
$T_{cl}$ is an inter-grain superconductive state.  In this paper we have
undertaken an extensive study on the magnetic and superconducting
properties of $M_{c}$Ta$_{2}$S$_{2}$C ($M$ = Fe, Co, Ni, Cu) from SQUID
(superconducting quantum interference device) DC and AC magnetic
susceptibility.  $M_{c}$Ta$_{2}$S$_{2}$C exhibits a variety of
superconducting and antiferromagnetic (AF) phase transitions, depending on
the kind of M. The magnetic properties of $M_{c}$Ta$_{2}$S$_{2}$C ($M$ =
Fe, Co) are mainly determined by magnetic behaviors of magnetic $M^{2+}$
ions in a crystal field such that the anion octahedra surrounding the
$M^{2+}$ ions are trigonally elongated along the $c$-axis.  In
Sec.~\ref{back} we present a simple review on the spin Hamiltonian of
Fe$^{2+}$ and Co$^{2+}$ under the trigonal crystal field.  We show that
both Fe$_{0.33}$Ta$_{2}$S$_{2}$C and Co$_{0.33}$Ta$_{2}$S$_{2}$C undergo an
AF phase transition at $T_{N}$ ($T_{N} = 117$ K for Fe and 18.6 K for Co),
reflecting the frustrated nature of the systems.  Like the pristine
Ta$_{2}$S$_{2}$C, $M_{c}$Ta$_{2}$S$_{2}$C ($M$ = Fe, Co, Ni, Cu) also
undergoes superconducting phase transitions at $T_{cu}$ and $T_{cl}$, where
$T_{cu}$ is dependent on the kind of $M^{2+}$ ions, while $T_{cl}$ (= 4.6
K) is common to all our systems.

\section{\label{back}BACKGROUND}
\subsection{\label{backA}Spin Hamiltonian of Fe$^{2+}$ in the trigonal
crystal field}

\begin{figure}
\includegraphics[width=8.5cm]{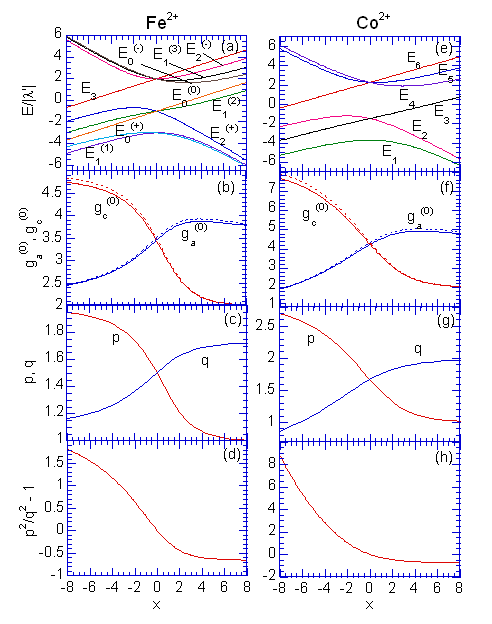}%
\caption{\label{fig01}(Color online) Derivation from Fe$^{2+}$ spin
Hamiltonian: (a) the energy levels, (b) $g$-factors $g_{c}^{(0)}$ and
$g_{a}^{(0)}$ with $k$ (= 0.9 (solid line) and 1 (dotted line)), (c) spin
anisotropy parameters $p$ and $q$, and (d) $p^{2}/q^{2}-1$, as a function
of $x$ ($= \delta_{0}/\lambda^{\prime}$).  $x = -1.27$ for
Fe$_{0.33}$Ta$_{2}$S$_{2}$C. Derivation from Co$^{2+}$ spin Hamiltonian:
(e) the energy levels, (f) $g$-factors $g_{c}^{(0)}$ and $g_{a}^{(0)}$ with
$k$ (= 0.9 (solid line) and 1 (dotted line)), (g) spin anisotropy
parameters $p$ and $q$, and (h) $p^{2}/q^{2} - 1$, as a function of $x$ ($=
\delta_{0}/\lambda^{\prime}$).  $x = 1.68$ for
Co$_{0.33}$Ta$_{2}$S$_{2}$C.}
\end{figure}

The free-ion 3d$^{6}$ $^{5}$D state of the Fe$^{2+}$ is split by the cubic
crystal field into the orbital doublet ($E$) and orbital triplet ($T_{2}$),
the latter being the lowest one.\cite{Kanamori1958,Inomata1967} We consider
the splitting of the orbital triplet by the perturbing Hamiltonian given by
\begin{equation}
H_{0} = -\lambda^{\prime}\textbf{l}\cdot\textbf{S}-\delta_{0}(l_{z}^{2}-2/3),
\label{eq01}
\end{equation}
where $\lambda^{\prime} =$ $k\lambda$ ($k$ $\approx$ 1 but less than unity)
and $S$ is the spin angular momentum of the magnitude 2.  A fictitious
angular momentum $\textbf{l}$ of the magnitude 1 is antiparallel to the
real orbital angular momentum $\textbf{L}$ ($= -k\textbf{l}$).  Figure
\ref{fig01}(a) shows the splitting of the ground orbital triplet by the
spin-orbit coupling $\lambda^{\prime}$ ($< 0$) and the trigonal field
$\delta_{0}$, where each energy level $E$ normalized by
$\vert\lambda^{\prime}\vert$ is plotted as a function of $x$ ($=
\delta_{0}/\lambda^{\prime}$).

All the energy states except for $E_{1}^{(1)}$ and $E_{0}^{(+)}$ might be
neglected, since these lowest levels lie 100 cm$^{-1}$ below the others. 
Thus we may use a fictitious spin $s$ = 1 for the lowest three states for
the singlet ($E = E_{0}^{(+)}$) and for the doublet ($E = E_{1}^{(1)}$) The
$g$-factors can be evaluated as $g_{c} = g_{c}^{(0)} + \Delta g$ and $g_{a}
= g_{a}^{(0)} + \Delta g$, where $\Delta g$ is due to the effect of
spin-orbit coupling in admixing the upper orbital levels into the ground
three orbitals.  Figure \ref{fig01}(b) shows the $g$-factors $g_{c}^{(0)}$
and $g_{a}^{(0)}$ as a function of $x$ with $k$ as a parameter ($k = 0.9$
and 1): $g_{c}^{(0)} > g_{a}^{(0)}$ for $x < 0$ and $g_{c}^{(0)} <
g_{a}^{(0)}$ for $x > 0$.  Note that $x = -1.27$ for FeCl$_{2}$.  If we
take the $z$ axis parallel to the $c$ axis, and $x$, $y$ axes perpendicular
to it, we have $S_{x} = qs_{x}$, $S_{y} = qs_{y}$, and $S_{z} = ps_{z}$. 
In Fig.  \ref{fig01}(c) we show the parameters $p$ and $q$ as a function of
$x$: $p>q$ for $x<0$ and $p<q$ for $x>0$.  The resultant spin Hamiltonian
for Fe$^{2+}$ is given by
\begin{equation}
H = - 2J \sum_{\langle i,j\rangle} \textbf{s}_{i}\cdot\textbf{s}_{j} 
 -D\sum_{i} (s_{iz}^{2}- 2/3)
 - 2J_{A}\sum_{\langle i,j\rangle} s_{iz}s_{jz},
\label{eq02}
\end{equation}
where $\langle i,j \rangle$ denotes the nearest neighbor pairs, $J =
q^{2}K$ and $K$ is the isotropic exchange energy with the form of
$-2KS_{i}\cdot S_{j}$ between the real spins $S_{i}$ and $S_{j}$,
$D\approx\delta_{0}/10$ ($> 0$) is the single ion anisotropy, and $J_{A}$
($=((p^{2}-q^{2})/q^{2})J$) (see Fig.~\ref{fig01}(d)) is the anisotropic
exchange interaction.  The spin anisotropy parameter $D_{eff}$ is defined
as $D_{eff}$ ($= D(s-1/2) + 2zsJ_{A}$).

\subsection{\label{backB}Spin Hamiltonian of Co$^{2+}$ in the trigonal
crystal field}
In a cubic crystal field the free-ion 3d$^{7}$ $^{4}$F state is split into
two orbital triplets and one orbital singlet with a triplet the
lowest.\cite{Kanamori1958,Lines1963,Oguchi1965} We consider the splitting
of the ground state orbital triplet $^{4}T_{1}$ into six Kramers doublets. 
The perturbing Hamiltonian consists of the spin-orbit coupling and trigonal
distortion of the crystal field,
\begin{equation}
H_{0}= -(3/2)\lambda^{\prime}\textbf{l}\cdot\textbf{S}-\delta_{0}(l_{z}^{2}-2/3),
\label{eq03}
\end{equation}
where $S$ is the spin angular momentum of the magnitude 3/2 and a
fictitious angular momentum $\textbf{l}$ of the magnitude 1 is antiparallel
to the real orbital angular momentum \textbf{$L$} ($= -3k\textbf{l}/2$). 
Figure \ref{fig01}(e) shows the energy level $E$ of the six Kramers
doublets normalized by $\vert\lambda^{\prime}\vert$, as a function of $x$
($= \delta_{0}/\lambda^{\prime}$).

For all values of $x$, $E_{\pm 1}$ is the lowest energy.  Since there are
only two states in this lowest Kramers doublet, the true spin $S$ (= 3/2)
can be replaced by a fictitious spin $s$ (= 1/2) within the ground state. 
The correction $\Delta$ $g$ is due to the effect of spin-orbit coupling in
admixing the upper orbital levels into the ground orbital triplet.  Figure
\ref{fig01}(f) shows the values of $g_{a}^{(0)}$ and $g_{c}^{(0)}$ as a
function of $x$ with $k$ as a parameter ($k = 0.9$ and 1.0).  Figure
\ref{fig01}(g) shows the parameters $p$ and $q$ as a function of $x$: $p>q$
for $x<0$ and $p<q$ for $x>0$.  The spin Hamiltonian of Co$^{2+}$ may be
written by Eq.(\ref{eq02}) without the second term.  The ratio $J_{A}/J$
($= (p^{2} - q^{2})/q^{2}$) (see Fig.~\ref{fig01}(h)) provides a measure
for the spin symmetry of the system.

\section{\label{exp}EXPERIMENTAL PROCEDURE}
Powdered samples of $M_{c}$Ta$_{2}$S$_{2}$C ($M$ = Fe, Co, Ni, Cu) were
prepared by Wally and Ueki,\cite{Wally1998c} where $c = 0.33$ for $M$ = Fe,
Co, $c$ = 0.25 for Ni, and 0.60 for Cu.  The detail of the synthesis and
structural characterization by x-ray diffraction was described by Wally and
Ueki.\cite{Wally1998c} The measurements of DC and AC magnetic
susceptibility were carried out using a SQUID magnetometer (Quantum Design
MPMS XL-5).  Before setting up samples at 298 K, a remnant magnetic field
was reduced to less than 3 mOe using an ultra-low field capability option. 
For convenience, hereafter this remnant field is noted as the state $H$ =
0.  The ZFC and FC susceptibility ($\chi_{ZFC}$ and $\chi_{FC}$) were
measured after a ZFC protocol, which consists of (i) cooling of the system
from 298 to 1.9 K at $H$ = 0 and (ii) applying $H$ at 1.9 K. The
susceptibility $\chi_{ZFC}$ was measured with increasing $T$ from 1.9 to
200 K. After annealing the system at 250 K (typically) for 1200 sec,
$\chi_{FC}$ was measured with decreasing $T$ from 200 to 1.9 K. The detail
of the measurements of DC and AC magnetic susceptibility is described in
Sec.~\ref{result}.

\section{\label{result}RESULT}
\subsection{\label{resultA}Paramagnetic susceptibility}

\begin{figure*}
\includegraphics[width=12.0cm]{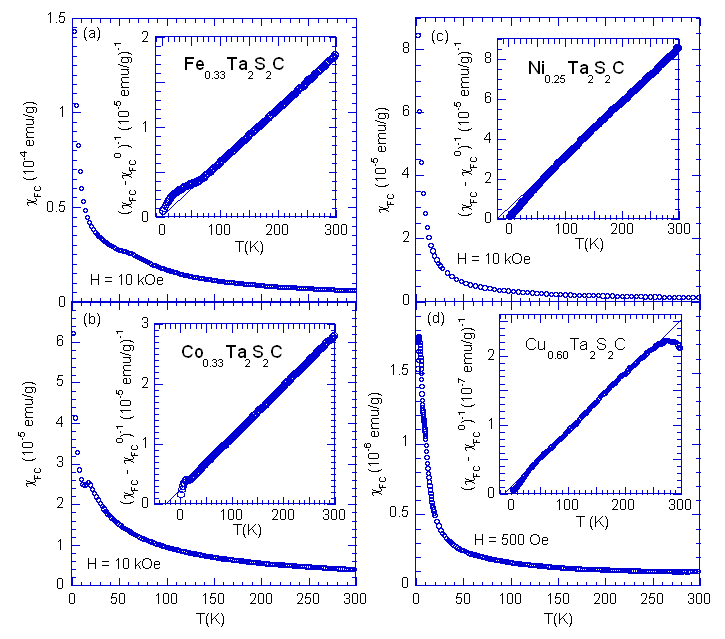}%
\caption{\label{fig02}(Color online) $T$ dependence of (a) $\chi_{FC}$ at
$H$ = 10 kOe for Fe$_{0.33}$Ta$_{2}$S$_{2}$C, (b) $\chi_{FC}$ at $H$ = 10
kOe for Co$_{0.33}$Ta$_{2}$S$_{2}$C, (c) $\chi_{FC}$ at $H$ = 10 kOe for
Ni$_{0.25}$Ta$_{2}$S$_{2}$C, and (d) $\chi_{FC}$ at $H$ = 500 Oe for
Cu$_{0.60}$Ta$_{2}$S$_{2}$C. The inset shows the $T$ dependence of the
reciprocal susceptibility ($\chi_{FC} - \chi_{FC}^{0})^{-1}$.  The solid
lines denote the least squares fitting curves to the Curie-Weiss law given
by Eq.(\ref{eq04}).  The fitting parameters are listed in Table
\ref{table1}.}
\end{figure*}

\begin{table*}
\caption{\label{table1}Magnetic susceptibility data of
$M_{c}$Ta$_{2}$S$_{2}$C ($M$ = Fe, Co, Ni, and Cu).  The least-squares
fitting parameters determined from the data of $\chi_{g}$ ($= \chi_{FC}$)
vs $T$ for $150 \leq T \leq 298$ K. The definition of the parameters is
given in the text.}
\begin{ruledtabular}
\begin{tabular}{lllll}
$M_{c}$ & $P_{eff}$($\mu_{B}$) & $\Theta$(K) & $\chi_{g}^{0}$ ($10^{-7}$ emu/g) & 
$\chi_{M}^{0}$ ($10^{-4}$ emu/M mole)\\
\hline
Fe$_{0.33}$ & 4.47 $\pm$ 0.03 &  -9.8 $\pm$ 1.6 & 0.91 $\pm$ 0.61 & 1.26 $\pm$ 0.84\\
Co$_{0.33}$ & 3.57 $\pm$ 0.04 & -26.3 $\pm$ 2.4 & 4.11 $\pm$ 0.48 & 5.70 $\pm$ 0.67\\
Ni$_{0.25}$ & 2.32 $\pm$ 0.03 & -19.7 $\pm$ 3.9 & 0.14 $\pm$ 0.28 & 0.26 $\pm$ 0.52\\
Cu$_{0.60}$ & 0.28 $\pm$ 0.01 & -13.7 $\pm$ 0.8 & 0.51 $\pm$ 0.06 & 0.40 $\pm$ 0.01\\
\end{tabular}
\end{ruledtabular}
\end{table*}

Figure \ref{fig02} shows the $T$ dependence of $\chi_{FC}$ at $H$ = 10 kOe
of $M_{c}$Ta$_{2}$S$_{2}$C ($M$ = Fe, Co, Ni) and at $H = 500$ Oe for $M$ =
Cu.  The susceptibility obeys a Curie-Weiss law,
\begin{equation}
\chi_{g} = \chi_{g}^{0} + C_{g}/(T - \Theta),
\label{eq04}
\end{equation}
where $\Theta$ (K) is the Curie-Weiss temperature, $C_{g}$ (emu K/g) is the
Curie-Weiss constant, and $\chi_{g}^{0}$ (emu/g) is the $T$-independent
susceptibility.  The gram susceptibilty $\chi_{g}$ of
$M_{c}$Ta$_{2}$S$_{2}$C is related to the molar susceptibility $\chi_{M}$
by $\chi_{M} = \zeta_{M}\chi_{g}$, where the molar mass is defined by
$\zeta_{M} = (M_{0}c + 438.039)$ (g/mole) and $M_{0}$ is the molar mass of
$M$ ion.  The molar Curie constant $C_{M}$ (emu K/M mole) [$=
\zeta_{M}C_{g} = (N_{A}\mu_{B}^{2}P_{eff}^{2})/3k_{B}$] is related to an
effective magnetic moment $P_{eff}$ by $C_{M} = P_{eff}^{2}/8$.  The
least-squares fits of the data of $\chi_{FC}$ vs $T$ for $150 \leq T \leq
298$ K to Eq.(\ref{eq04}) yield the parameters ($\Theta$, $C_{g}$, and
$\chi_{g}^{0}$) listed in Table \ref{table1}.

\begin{table}
\caption{\label{table2}Parameters of spin Hamiltonian for
$M_{c}$Ta$_{2}$S$_{2}$C. $s$ is a fictitious spin and $S$ is a spin
predicted from the Hund rule.  $J$ is the intraplanar exchange
interaction.}
\begin{ruledtabular}
\begin{tabular}{lllll}
$M_{c}$ & $s$ & $S$ & $g$ & $J$ (K)\\
\hline
Fe$_{0.33}$ & 1 & 2 & 3.16 $\pm$ 0.02 & -1.22 $\pm$ 0.20\\
Co$_{0.33}$ & 1/2 & 3/2 & 4.12 $\pm$ 0.05 & -8.75 $\pm$ 0.78\\
Ni$_{0.25}$ & \_ & 1 & 1.64 $\pm$ 0.02 & -2.46 $\pm$ 0.49 \\
Cu$_{0.60}$ & \_ & 1/2 & 0.324 $\pm$ 0.003 & -4.57 $\pm$ 0.27\\
\end{tabular}
\end{ruledtabular}
\end{table}

We find that $\Theta$ is negative for $M_{c}$Ta$_{2}$S$_{2}$C with M =
Fe$^{2+}$, Co$^{2+}$, Ni$^{2+}$, and Cu$^{2+}$ ions, suggesting the AF
nature of the intraplanar exchange interaction between $M^{2+}$ spins. 
$M_{c}$Ta$_{2}$S$_{2}$C magnetically behaves like a quasi two-dimensional
(2D) AF. Structurally the interplanar distance ($\approx 8.6 \AA$) between
$M$ ions in the adjacent $M$ layers is longer than the intraplanar distance
($\sqrt{3}a_{0} = 5.72 \AA$) for $c = 1/3$ between $M^{2+}$ ions in the
same $M$ layer.  The existence of S-Ta-C-Ta-S sandwiched layered structures
greatly reduces the interplanar exchange interaction between $M^{2+}$ ions
in the adjacent vdw gaps.  Then the intraplanar exchanage interaction $J$
can be estimated using the relation $\Theta = 2zJs(s+1)/3k_{B}$, where $z$
(= 6) is the number of the nearest neighbor $M$ atoms.  The values of $J$
and spin $s$ for $M_{c}$Ta$_{2}$S$_{2}$C are listed in Table \ref{table2},
where $s = S = 1$ for Ni$^{2+}$ and 1/2 for Cu$^{2+}$.

The $T$-independent gram-susceptibility ($\chi_{g}^{0}$) of
$M_{c}$Ta$_{2}$S$_{2}$C is positive.  The paramagnetic nature of
$M_{c}$Ta$_{2}$S$_{2}$C arises from magnetic $M^{2+}$ ions.  This is in
contrast to the diamagnetic nature of the pristine Ta$_{2}$S$_{2}$C. Note
that $\chi_{FC}$ of Ta$_{2}$S$_{2}$C takes a negative constant (diamagnetic
susceptibility) above 150 K: $\chi_{g}^{0}$(Ta$_{2}$S$_{2}$C) $= (-1.44 \pm
0.01) \times 10^{-7}$ emu/g at 298 K. The $T$-independent molar
susceptibility ($\chi_{M}^{0}$) of $M_{c}$Ta$_{2}$S$_{2}$C, as listed in
Table \ref{table2}, is related to $\chi_{g}^{0}$ as $\chi_{M}^{0} =
\zeta_{M} \chi_{g}^{0}$.  The value of $\chi_{M}^{0}$ for $M$ = Co is much
larger than that that for $M$ = Fe, Ni, and Cu.  The diamagnetic
contribution from Ta$_{2}$S$_{2}$C is subtracted from $\chi_{g}^{0}$ for
Co$_{0.33}$Ta$_{2}$S$_{2}$C. The effective molar susceptibility
($\chi_{M}^{0}$)$_{c}$ of Co$_{0.33}$Ta$_{2}$S$_{2}$C can be calculated as
$(7.61 \pm 0.68) \times 10^{-4}$ emu/Co mole) using the relation
\[
(\chi_{M}^{0})_{c}=[m\chi_{g}^{0}(Co_{x}Ta_{2}S_{2}C) - m_{0}
\chi_{g}^{0}(Ta_{2}S_{2}C)]/c,
\]
with $c = 0.33$, where $m$ and $m_{0}$ is the molar mass of
Co$_{c}$Ta$_{2}$S$_{2}$C and Ta$_{2}$S$_{2}$C, respectively.  Such a large
value of ($\chi_{M}^{0})_{c}$ may be partly due to the Van Vleck
susceptibility of Co$^{2+}$ spin arising from the narrow energy difference
between the ground state with $E_{1}$ and five Kramers doublets with
$E_{2}$ - $E_{6}$.\cite{Lines1963} For $x = 1.68$ and $k = 0.9$ for
Co$^{2+}$ spin, the Van Vleck susceptibility can be calculated as
$\chi_{V}^{z} = 4.19N_{A}\mu_{B}^{2}/(\vert\lambda\vert)$ and $\chi_{V}^{x}
= 7.29N_{A}\mu_{B}^{2}/(\vert\lambda\vert)$, where $N_{A}$ is the Avogadro
number.  The Van Vleck susceptibility for the powdered system [$=
(2\chi_{V}^{x} + \chi_{V}^{z})/3$] is estimated as $9.1 \times 10^{-3}$
emu/Co mole.  Thus our value of $(\chi_{M}^{0})_{c}$ cannot be well
explained only from this model.

\subsection{\label{resultB}$\chi_{ZFC}$ and $\chi_{FC}$ at $H = 1$ Oe}

\begin{figure*}
\includegraphics[width=12.0cm]{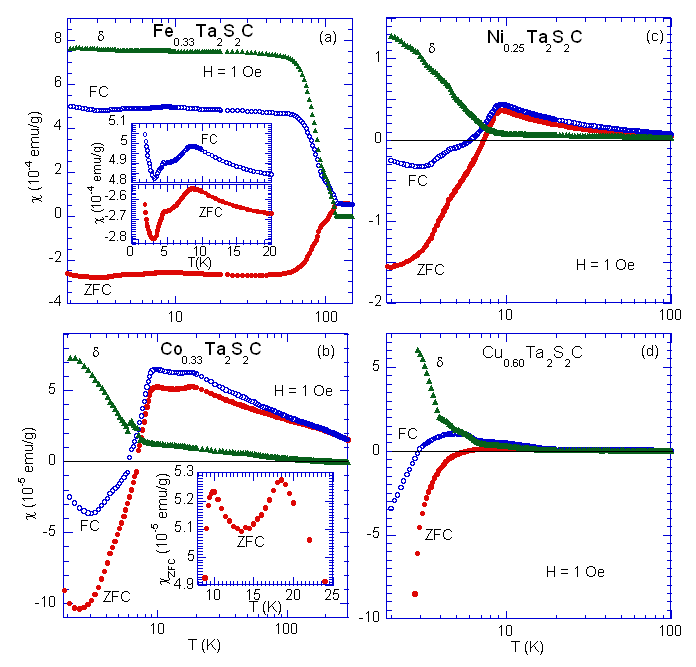}%
\caption{\label{fig03}(Color online) $T$ dependence of $\chi_{ZFC}$,
$\chi_{FC}$, and $\delta$ ($= \chi_{FC}-\chi_{ZFC}$) for (a)
Fe$_{0.33}$Ta$_{2}$S$_{2}$C, (b) Co$_{0.33}$Ta$_{2}$S$_{2}$C, (c)
Ni$_{0.25}$Ta$_{2}$S$_{2}$C, and (d) Cu$_{0.60}$Ta$_{2}$S$_{2}$C. $H$ = 1
Oe.  The units of (c) and (d) are the same as those of (a) and (b),
respectively.}
\end{figure*}

We have measured the $T$ dependence of $\chi_{ZFC}$ and $\chi_{FC}$ at $H =
1$ Oe for $M_{c}$Ta$_{2}$S$_{2}$C. Figure \ref{fig03}(a) shows the $T$
dependence of $\chi_{ZFC}$, $\chi_{FC}$, and $\delta$ ($= \chi_{FC} -
\chi_{ZFC}$) for Fe$_{0.33}$Ta$_{2}$S$_{2}$C. Both $\chi_{ZFC}$ and
$\chi_{FC}$ have a kink at 4.5 K and a peak at 8.7 K. The sign of
$\chi_{ZFC}$ at $H = 1$ Oe is negative below 92 K. Note that for $H \geq
10$ Oe, the sign of $\chi_{ZFC}$ is positive at any $T$ (see
Fig.~\ref{fig04}).  The susceptibility $\chi_{FC}$, whose sign is positive
at any $T$ and $H$, exhibits a drastic increases below 117 K with
decreasing $T$ and almost saturates below 60 K.

The deviation $\delta$ drastically increases with decreasing $T$ below 117
K, suggesting the frustrated nature of the system (the 2D antiferromagnet
on the triangular lattice).  The negative sign of $\chi_{ZFC}$ below 92 K
at $H$ = 1 Oe may be related to the diamagnetic susceptibility of the
pristine Ta$_{2}$S$_{2}$C as a host,\cite{Walter2004} which is a feature common to
systems having 1T-TaS$_{2}$ type structure.\cite{DiSalvo1980} We note that
the value of $\chi_{ZFC}$ at $H$ = 1 Oe and $T = 1.9$ K is $-2.5 \times
10^{-4}$ emu/g for Fe$_{0.33}$Ta$_{2}$S$_{2}$C, which is in contrast to
that ($\chi_{ZFC} = -1.7 \times 10^{-3}$ emu/g) for Ta$_{2}$S$_{2}$C as a
type II superconductor.

Figure \ref{fig03}(b) shows the $T$ dependence of $\chi_{ZFC}$,
$\chi_{FC}$, and $\delta$ at $H = 1$ Oe for Co$_{0.33}$Ta$_{2}$S$_{2}$C.
The susceptibility $\chi_{ZFC}$ has peaks at 10.0 and 18.6 K, while
$\chi_{FC}$ has peaks at 9.8 K and 18.6 K. The peaks at 10 K and 18.6 K are
related to the superconductivity ($T_{cu} = 9.1$ K) and the AF spin
ordering ($T_{N} = 18.6$ K).  Both $\chi_{ZFC}$ and $\chi_{FC}$ are
negative at low $T$, while the deviation $\delta$, which is positive at any
$T$, decreases drastically with increasing $T$ and reduces to zero at
higher $T$.  This system undergoes a superconducting transition at $T_{cu}$
at which $\delta$ tends to reduce to zero.

Figure \ref{fig03}(c) shows the $T$ dependence of $\chi_{ZFC}$,
$\chi_{FC}$, and $\delta$ at $H = 1$ Oe for Ni$_{0.25}$Ta$_{2}$S$_{2}$C.
The susceptibility $\chi_{FC}$ at $H = 1$ Oe shows a local minimum at 3.0 K
and a broad peak at 9.40 K, while $\chi_{ZFC}$ shows a peak at 9.40 K. The
deviation $\delta$ exhibits a step-like change at 9.85 K. This system
undergoes a superconducting transition at $T_{cu}$ ($= 8.69 \pm 0.04$ K),
at which the data of d$\delta$/d$T$ vs $T$ has a local minimum (see
Sec.~\ref{resultE}).

Figure \ref{fig03}(d) shows the $T$ dependence of $\chi_{ZFC}$,
$\chi_{FC}$, and $\delta$ at $H = 1$ Oe for Cu$_{0.60}$Ta$_{2}$S$_{2}$C.
The susceptibility $\chi_{FC}$ shows a broad peak at 4.9 K, while
$\chi_{ZFC}$ shows a broad peak around 8 - 9 K. The deviation $\delta$
exhibits three step-like changes around 2.9, 3.9 and 6.5 K. This system
undergoes a superconducting transition at $T_{cu} = 6.4$ K, at which the
data of d$\delta$/d$T$ vs $T$ has a local minimum.

\subsection{\label{resultC}Fe$_{0.33}$Ta$_{2}$S$_{2}$C}

\begin{figure}
\includegraphics[width=7.0cm]{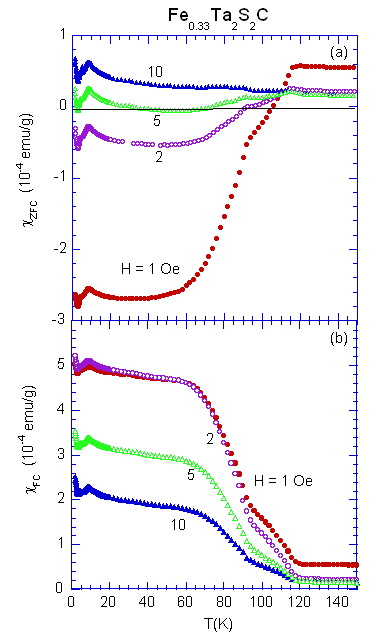}%
\caption{\label{fig04}(Color online) $T$ dependence of (a) $\chi_{ZFC}$ and
(b) $\chi_{FC}$ at $H$ (= 1, 2, 5, and 10 Oe) for
Fe$_{0.33}$Ta$_{2}$S$_{2}$C.}
\end{figure}

\begin{figure}
\includegraphics[width=8.5cm]{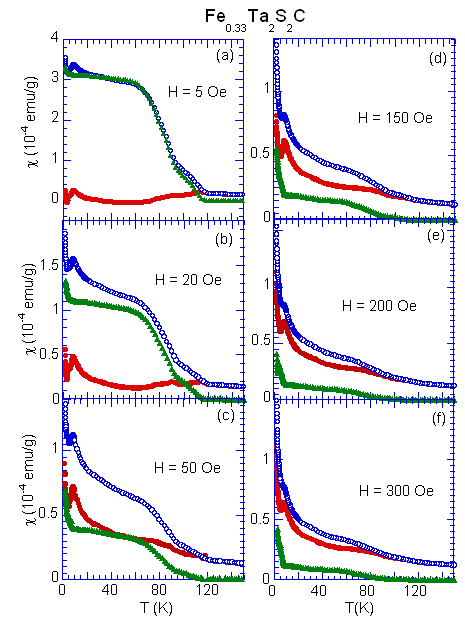}%
\caption{\label{fig05}(Color online) (a)-(f) $T$ dependence of
$\chi_{ZFC}$, $\chi_{FC}$, and $\delta$ at various $H$ for
Fe$_{0.33}$Ta$_{2}$S$_{2}$C. The units of (d)-(f) are the same as those of
(a)-(c).}
\end{figure}

In Figs.~\ref{fig04}(a) and (b) we show the $T$ dependence of $\chi_{ZFC}$
and $\chi_{FC}$ at $H$ (= 1, 2, 5, and 10 Oe) for
Fe$_{0.33}$Ta$_{2}$S$_{2}$C, respectively.  In Fig.5 we show the $T$
dependence of $\chi_{ZFC}$, $\chi_{FC}$, and $\delta$ at low $H$ for
Fe$_{0.33}$Ta$_{2}$S$_{2}$C. The susceptibility $\chi_{ZFC}$ at $H = 1$ and
2 Oe is negative below 106 K and 93 K, respectively.  In contrast,
$\chi_{ZFC}$ above 10 Oe and $\chi_{FC}$ at any $H$ are positive at any
$T$.  The susceptibility $\chi_{ZFC}$ at $H = 1$ Oe shows two peaks at
$T_{cu}$ (= 8.8 K) and $T_{N}$ (= 117 K), and two cusps at $T_{cl}$ (= 4.6
K) and $T_{N}^{\prime}$ ($\approx 94$ K).  The susceptibility $\chi_{FC}$
(typically at 20 Oe) starts to increase with decreasing $T$ around $T_{N}$. 
It shows a peak at $T_{cu}$, and a cusp at $T_{cl}$.  The deviation
$\delta$ at $H = 20$ Oe, for example, undergoes step-like increases around
$T_{N}$ and $T_{cu}$, indicating that the irreversible effect of
magnetization occurs.  At $H = 1200$ Oe there is no appreciable difference
between $\chi_{ZFC}$ and $\chi_{FC}$ at any $T$.  Note that $\delta$ is
almost equal to zero above $T_{N}$.  The deviation $\delta$ (typically at
$H = 5$ and 20 Oe) shows step-like changes at $T_{N}$, $T_{N}^{\prime}$,
and $T_{cu}$, with decreasing $T$.

\begin{figure}
\includegraphics[width=7.0cm]{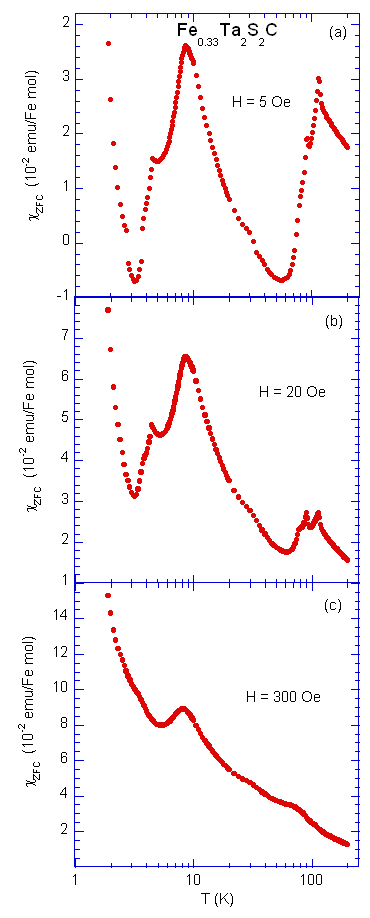}%
\caption{\label{fig06}(Color online) (a)-(c) $T$ dependence of $\chi_{ZFC}$
(emu/Fe mole) at $H$ (= 5, 20, and 300 Oe) for
Fe$_{0.33}$Ta$_{2}$S$_{2}$C.}
\end{figure}

Figure \ref{fig06} shows the $T$ dependence of $\chi_{ZFC}$ (emu/Fe mole)
at $H = 5$, 20, and 300 Oe, where $\chi_{ZFC}$(emu/Fe mole) =
$\zeta_{Fe}\chi_{ZFC}$ (emu/g) with $\zeta_{Fe} = 1383.23$ g/Fe mole.  The
susceptibility $\chi_{ZFC}$ at $H = 20$ Oe exhibits peaks around $T_{cl}$,
$T_{cu}$, $T_{N}^{\prime}$, and $T_{N}$, while $\chi_{ZFC}$ at $H = 300$ Oe
shows a broad peak around $T_{cu}$.  The peak values of $\chi_{ZFC}$ are
0.036 emu/Fe mole at $T_{cu}$ and 0.03 emu/Fe mole at $T_{N}$,
respectively.  Note that the $T$ dependence of the DC susceptibility $\chi$
of Fe$_{c}$Nb$_{2}$S$_{2}$C ($c = 0.4$, 0.5, and 0.6) between 80 and 550 K
has been reported by Boller and Hiebl.\cite{Boller1992} The susceptibility
$\chi$ exhibits a cusp-like behavior at a peak temperature $T_{p}$, which
decreases with decreasing Fe concentration $c$: $T_{p} = 346$ K at $c =
0.6$ and $T_{p} = 180$ K at $c = 0.4$.  An extrapolation of the possible
linear relation of $T_{p}$ vs $c$ may lead to $T_{p} = 137$ K at $c =
0.33$, suggesting that a possible peak temperature $T_{p}$ in
Fe$_{0.33}$Nb$_{2}$S$_{2}$C may correspond to $T_{N}$ in
Fe$_{0.33}$Ta$_{2}$S$_{2}$C. The magnetic neutron powder diffraction of
Fe$_{0.5}$Nb$_{2}$S$_{2}$C shows no evidence for the possible long range
spin order below $T_{p}$ = 305 K (Boller).\cite{Boller2003}

\begin{figure}
\includegraphics[width=7.0cm]{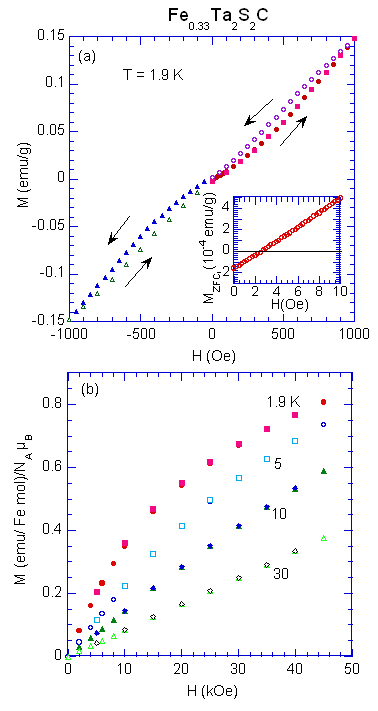}%
\caption{\label{fig07}(Color online) (a) $H$ dependence of $M$ at $T = 1.9$
K for Fe$_{0.33}$Ta$_{2}$S$_{2}$C. The measurement of $M$ was made by
changing $H$ in sequence from 0 to 1 kOe ($M$ coincides with $M_{ZFC}$),
from 1 kOe to -1 kOe, and from -1 kOe to 1 kOe, after the sample was cooled
from 298 K to 1.9 K in the absence of $H$.  The inset shows the $H$
dependence of $M_{ZFC}$ at $T = 1.9$ K.  (b) $H$ dependence of $M_{FC}$ at
various $T$ for Fe$_{0.33}$Ta$_{2}$S$_{2}$C.}
\end{figure}

In Fig.~\ref{fig07}(a) we show the $H$ dependence of $M$ at $T = 1.9$ K for
Fe$_{0.33}$Ta$_{2}$S$_{2}$C. The magnetizastion $M$ was measured by
changing $H$ in sequence from 0 to 1 kOe, from 1 kOe to -1 kOe, and from -1
kOe to 1 kOe, after the sample was cooled from 298 K to 1.9 K in the
absence of $H$.  The magnetization $M$ slightly shows a magnetic
hysteresis.  The inset of Fig.~\ref{fig07}(a) shows the $H$ dependence of
$M_{ZFC}$ at $T$ = 1.9 K, where the measurement was carried out after the
ZFC cooling from 298 to 1.9 K in the absence of $H$.  The value of
$M_{ZFC}$ is negative at $H = 0$ and increases with increasing $H$.  The
sign of $M_{ZFC}$ changes around $H = 2.6$ Oe from negative to positive. 
It increases with further increasing $H$.

Figure \ref{fig07}(b) shows the $H$ dependence of the $M_{FC}$ (in the
units of emu/Fe mole) divided by $N_{A}\mu_{B}$ (= 5584.94 emu/Fe mole) at
various $T$ for Fe$_{0.33}$Ta$_{2}$S$_{2}$C. No appreciable hysteresis
effect is observed: the $M$-$H$ curve with increasing $H$ is almost the
same as that with decreasing $H$.  The saturation magnetization $M_{s}$ is
equal to $N_{A}\mu_{B}gs$ (= 17648.4 emu/Fe mole), when $g$ ($= 3.16 \pm
0.02$) and the fictitious spin ($s = 1$) for Fe$^{2+}$ are assumed.  The
magnetization $M_{FC}$ reaches $4.56 \times 10^{3}$ emu/Fe mole at $H = 45$
kOe and $T = 1.9$ K, corresponding to $M/M_{s} = 0.26$, which is close to
1/3 corresponding to the metamagnetic plateau in the $M$-$H$ curve observed
in quasi 2D $XY$-like AF on the triangular lattice C$_{6}$Eu (Eu$^{2+}$
with $S = 7/2$).\cite{Suematsu1981,Date1983,Sakakibara1984}

\begin{figure*}
\includegraphics[width=12.0cm]{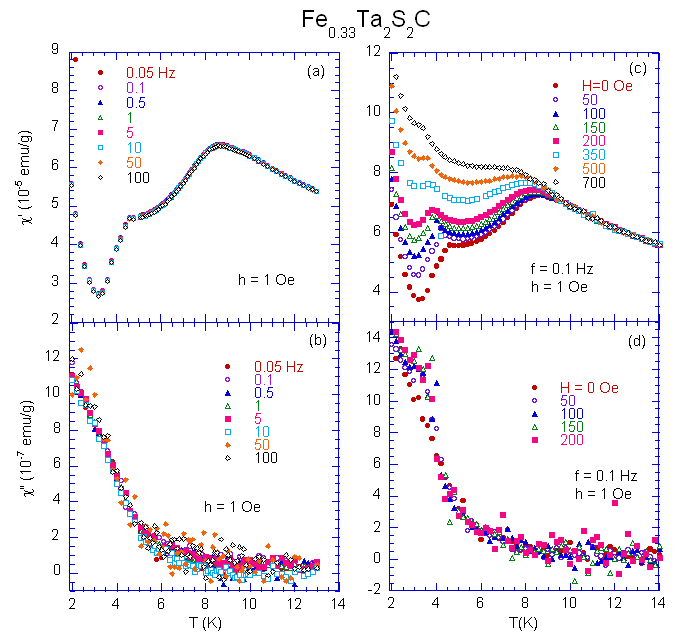}%
\caption{\label{fig08}(Color online) $T$ dependence of (a) $\chi^{\prime}$
and (b) $\chi^{\prime\prime}$ at $f$ (= 0.05 - 100 Hz) for
Fe$_{0.33}$Ta$_{2}$S$_{2}$C. $h$ = 1 Oe.  $T$ dependence of (c)
$\chi^{\prime}$ and (d) $\chi^{\prime\prime}$ at various $H$ for
Fe$_{0.33}$Ta$_{2}$S$_{2}$C. $f$ = 0.1 Hz.  $h$ = 1 Oe.  The units of (c)
and (d) are the same as those of (a) and (b), respectively.  Note that the
background part of $\chi^{\prime}$ and $\chi^{\prime\prime}$ at $f$ = 0.1
Hz and $h$ = 1 Oe in (a) and (b) is different from those in (c) and (d). 
These two measurements were carried out independently.  The method of the
measurements is different: (a) and (b) for the $f$ scan at each fixed $T$
and (c) and (d) for the $T$ scan at $f$ = 0.1 Hz as $T$ was increased at
each $H$ (after each $T$ scan, $H$ was increased at 14 K and then $T$ was
decreased to 1.9 K in the presence of $H$).}
\end{figure*}

Figures \ref{fig08}(a) and (b) show the $T$ dependence of the dispersion
$\chi^{\prime}$ and the absorption $\chi^{\prime\prime}$ in the absence of
$H$ where the amplitude of the AC field is $h = 1$ Oe and the AC frequency
$f$ is between 0.05 Hz and 100 Hz.  After the sample was cooled from 298 to
1.9 K at $H = 0$, the AC susceptibility was measured at fixed $T$ as a
function of $f$ during the process of increasing $T$ from 1.9 to 13 K. The
dispersion $\chi^{\prime}$ shows a cusp at $T_{cl}$ and peak around
$T_{cu}$, while $\chi^{\prime\prime}$ decreases with increasing $T$ and
reduces to zero above $T_{cu}$.  No frequency dependence of $\chi^{\prime}$
and $\chi^{\prime\prime}$ is observed.  Figures \ref{fig08}(c) and (d) show
the $T$ dependence of $\chi^{\prime}$ and $\chi^{\prime\prime}$ in the
presence of $H$, where $f$ = 0.1 Hz and $h = 1$ Oe.  The shift of the peak
at $T_{cu}$ to the low-$T$ side with increasing $H$.  The absorption
$\chi^{\prime\prime}$ is slightly dependent on $H$ below $T_{cl}(H)$. 

\begin{figure}
\includegraphics[width=7.0cm]{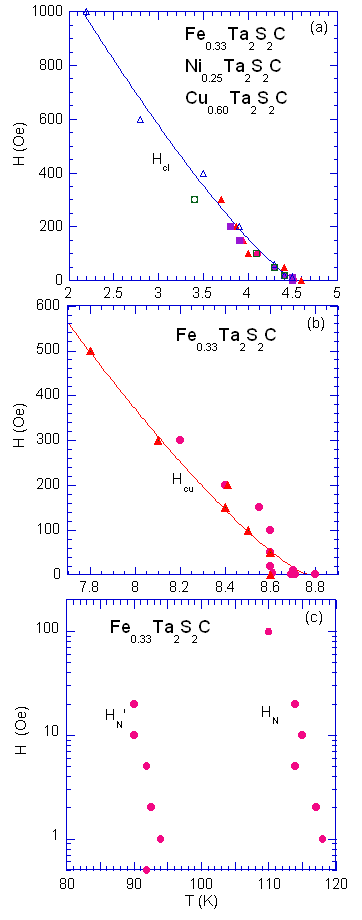}%
\caption{\label{fig09}(Color online) $H$-$T$ diagram of
$M_{c}$Ta$_{2}$S$_{2}$C. (a) The line $H_{cl}(T)$ for
Fe$_{0.33}$Ta$_{2}$S$_{2}$C [peak temperatures of $\chi_{ZFC}$ vs $T$
({\Large $\bullet$}), $\chi_{FC}$ vs $T$ ($\blacksquare$), and
$\chi^{\prime}$ vs $T$ ($\blacktriangle$) as a function of $H$],
Ni$_{0.25}$Ta$_{2}$S$_{2}$C [peak temperatures of $\chi_{ZFC}$ vs $T$
({\Large $\circ$}) and $\chi_{FC}$ vs $T$ ($\square$) as a function of $H$],
and Cu$_{0.60}$Ta$_{2}$S$_{2}$C [peak temperatures of $\chi_{FC}$ vs $T$
($\Delta$) as a function of $H$].  (b) The line $H_{cu}(T)$ for
Fe$_{0.33}$Ta$_{2}$S$_{2}$C: peak temperatures of $\chi_{ZFC}$ vs $T$
({\Large $\bullet$}) and $\chi^{\prime}$ vs $T$ ($\blacktriangle$) as a
function of $H$.  (c) The lines $H_{N}^{\prime}(T)$ and $H_{N}(T)$ for
Fe$_{0.33}$Ta$_{2}$S$_{2}$C: peak temperatures of $\chi_{ZFC}$ vs $T$
({\Large $\bullet$}) as a function of $H$.  The solid lines of (a) and (b)
denote least-squares fitting curves of the data to the power law form
Eq.(\ref{eq05}).}
\end{figure}

In Fig.~\ref{fig09} we show the $H$-$T$ diagram for
Fe$_{0.33}$Ta$_{2}$S$_{2}$C, where the temperatures of peaks or cusps of
$\chi_{ZFC}$ vs $T$ and $\chi^{\prime}$ vs $T$ at various $H$ are plotted
as a function of $H$.  There are four lines denoted by $H_{c2}^{(l)}(T)$,
$H_{c2}^{(u)}(T)$, $H_{N}^{\prime}(T)$, and $H_{N}(T)$ which correspond to
$T_{cl}(H)$, $T_{cu}(H)$, $T_{N}^{\prime}(H)$, and $T_{N}(H)$,
respectively.  The two lower temperatures ($T_{cu}$ and $T_{cl}$) are
related to the onset of superconductivity in the sample in the pristine
Ta$_{2}$S$_{2}$C sample.  The phase transition at $T_{N}$ (= 117 K), where
the ZFC and FC curves split, is most likely associated with an AF magnetic
ordering of the spin system.  The possible transition at 
$T_{N}^{\prime}$ (= 94 K) may be
assigned to some change of this magnetic structure.  The magnetic ordering
occurs within the intercalated magnetic layers and does not destroy the
superconducting transition that occurs at lower temperature.

The ratio $\eta$ ($=T_{N}/\vert\Theta\vert$) provides a measure for the
degree of the deviation from a molecular field theory.  It is known that
the ratio $\eta$ is smaller than 1 for conventional spin systems and tends
to decrease with the change of the dimension $d$ from 3 to 2, typically
$T_{N}/\vert\Theta\vert = 0.49$ for
FeCl$_{2}$,\cite{Birgeneau1972,Starr1940} 0.65 for
CoCl$_{2}$,\cite{Starr1940,Hutchings1973} and 0.77 for
NiCl$_{2}$.\cite{Starr1940,Lindgard1975} It is unexpected that AF magnetic
phase transition occurs at such a high temperature as 117 K, when the
derived $\Theta$ is only -9.8 K. However, a Curie-Weiss temperature
$\Theta$ derived from a limited temperature window does not tell all about
the interactions in the system.  The Fe$^{2+}$ ions from an triangular
lattice with the side having the in-plane distance $5.72 \AA$ for
Fe$_{0.33}$Ta$_{2}$S$_{2}$C. Since the $c$-axis stacking sequence is
Ta-C-Ta-S-Fe-S-Ta, the interaction between Fe$^{2+}$ spins is considered to
be a superexchange interaction between nearest neighbor Fe$^{2+}$ spins
through a superexchange path Fe-S-Fe.  This exchange interaction is roughly
estimated as $\vert J\vert = 15$ K if $T_{N}$ is equal to
$\vert\Theta\vert$ ($= 8 \vert J\vert$).  At present we have no explanation
for the origin of such a large $\vert J\vert$.

\subsection{\label{resultD}Co$_{0.33}$Ta$_{2}$S$_{2}$C}

\begin{figure}
\includegraphics[width=7.0cm]{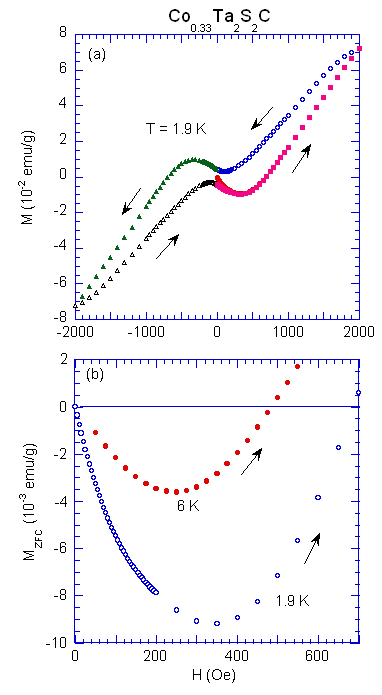}%
\caption{\label{fig10}(Color online) (a) Magnetization curve ($M$ vs $H$)
at $T = 1.9$ K for Co$_{0.33}$Ta$_{2}$S$_{2}$C. (b) Detail of $M_{ZFC}$ vs
$H$ at $T = 1.9$ and 6.0 K.}
\end{figure}

Figure \ref{fig10}(a) shows the hysteresis loop of the magnetization $M$ at
$T = 1.9$ K. After the sample was quenched from 298 to 1.9 K at $H = 0$,
the measurement was carried out with varying $H$ from 0 to 2 kOe at $T$,
from $H = 2$ to -2 kOe, and from $H = -2$ to 2 kOe.  The $M$-$H$ curve at
1.9 K shows a large hysteresis and a remnant magnetization.  Structural
imperfections or defect in the sample may play an role of flux pinning,
resulting in a inhomogeneous type-II superconductor.  Figure \ref{fig10}(b)
shows typical data of $M_{ZFC}$ vs $H$ at various $T$.  Before each
measurement, the sample was kept at 20 K at $H = 0$ for 1200 sec and then
it was quenched from 20 K to $T$ ($<4$ K).  The magnetization $M_{ZFC}$ at
$T$ was measured with increasing $H$ ($0 \leq H \leq 700$ Oe).  The
magnetization $M_{ZFC}$ exhibits a single local minimum at a characteristic
field for $T<T_{cu}$, shifting to the low-$H$ side with increasing $T$. 
The local minimum field at 1.9 K ($\approx 350$ Oe) for
Co$_{0.33}$Ta$_{2}$S$_{2}$C is much higher than that at 1.9 K for the
pristine Ta$_{2}$S$_{2}$C (= 60 Oe).  The lower critical field $H_{c1}(T)$
is defined not as the first minimum point of the $M_{ZFC}$ vs $H$, but as
the first deviation point from the linear portion due to the penetration of
magnetic flux into the sample: $H_{c1}$($T=1.9$ K) $= 65 \pm 5$ Oe.  Using
a conventional relation\cite{Ketterson1999} $H_{c1}(T) = H_{c1}(0)[1-
(T/T_{cu})^{2}]$ with $T_{cu} = 9.14 \pm 0.03$ K, $H_{c1}(0)$ can be
estimated as $68 \pm 5$ Oe.  Note that no demagnetization factor is taken
into account in deriving $H_{c1}(T)$.

\begin{figure*}
\includegraphics[width=12.0cm]{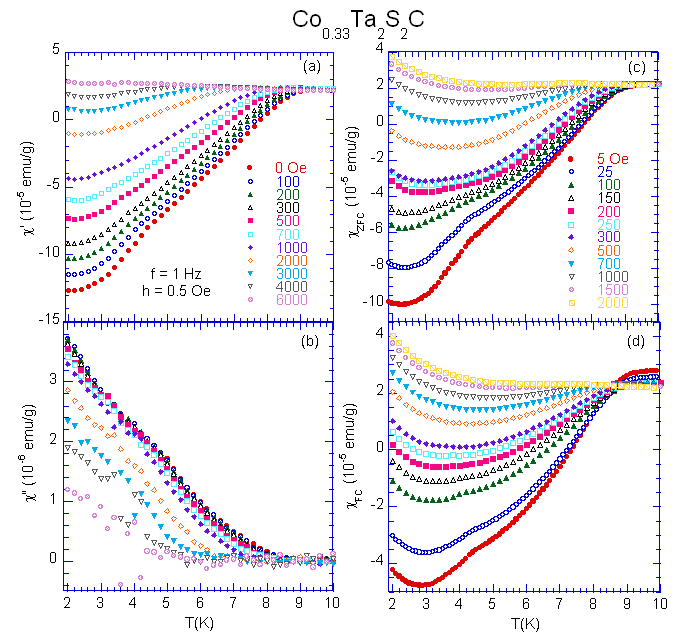}%
\caption{\label{fig11}(Color online) $T$ dependence of (a) $\chi^{\prime}$
and (b) $\chi^{\prime\prime}$ at various $H$ for
Co$_{0.33}$Ta$_{2}$S$_{2}$C. $f = 1$ Hz.  $h = 0.5$ Oe.  $T$ dependence of
(c) $\chi_{ZFC}$ and (d) $\chi_{FC}$ at various $H$ for
Co$_{0.33}$Ta$_{2}$S$_{2}$C.}
\end{figure*}

\begin{figure}
\includegraphics[width=7.0cm]{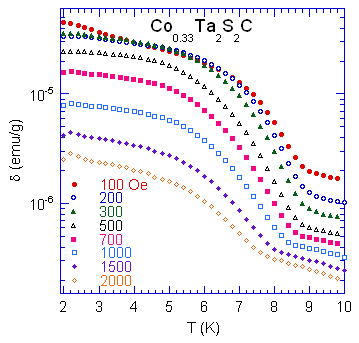}%
\caption{\label{fig12}(Color online) $T$ dependence of $\delta$ for
Co$_{0.33}$Ta$_{2}$S$_{2}$C.}
\end{figure}

\begin{figure}
\includegraphics[width=7.0cm]{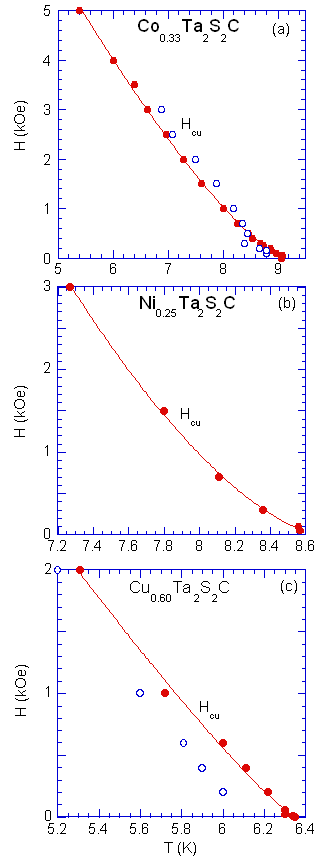}%
\caption{\label{fig13}(Color online) (a) $H$-$T$ diagram for
Co$_{0.33}$Ta$_{2}$S$_{2}$C: $H_{cu}(T)$ determined from the data of
$\chi^{\prime}$ vs $T$ (closed circles) and $\delta$ vs $T$ (open circles)
with $H$.  (b) $H$-$T$ phase diagram for Ni$_{0.25}$Ta$_{2}$S$_{2}$C:
$H_{cu}(T)$ determined from the data of $\delta$ vs $T$ with $H$.  (c)
$H$-$T$ phase diagram for Cu$_{0.60}$Ta$_{2}$S$_{2}$C: $H_{cu}(T)$
determined from the data of d$\delta$/d$T$ vs $T$ (closed circle) and
d$\chi_{ZFC}$/d$T$ vs $T$ (open circle) with $H$.  The solid lines denote
least-squares fitting curves to the power law form Eq.(\ref{eq05}).}
\end{figure}

Figures \ref{fig11}(a) and (b) show the $T$ dependence of $\chi^{\prime}$
and $\chi^{\prime\prime}$ for Co$_{0.33}$Ta$_{2}$S$_{2}$C, where $f$ = 1 Hz
and $h$ = 0.5 Oe.  The $T$ dependence of $\chi^{\prime}$ and
$\chi^{\prime\prime}$ is strongly dependent on $H$.  Below the critical
temperature $T_{cu}(H)$, the sign of $\chi^{\prime}$ and
$\chi^{\prime\prime}$ is negative and positive, respectively, suggesting
the existence of the Meissner effect.

Figures \ref{fig11}(c) and (d) show the $T$ dependence of $\chi_{ZFC}$ and
$\chi_{FC}$ for Co$_{0.33}$Ta$_{2}$S$_{2}$C at various $H$.  In
Fig.~\ref{fig12}, we show the $T$ dependence of $\delta$ for
Co$_{0.33}$Ta$_{2}$S$_{2}$C. In Fig.~\ref{fig13}(a) we show the plot of
$T_{cu}(H)$ as a function of $H$, where $T_{cu}(H)$ is defined as a
temperature at which $\chi^{\prime}$ at $H$ coincides with that at $H$ = 7
kOe.  We find that $T_{cu}(H)$ is almost equal to a temperature at which
$\delta$ exhibits a kink just after a drastic decrease in $\delta$ with
increasing $T$.  An upward curvature in $H_{c2}^{(u)}$ vs $T$ indicates
that $H_{c2}^{(u)}(T)$ can be described by
\begin{equation}
H_{c2}^{(u)}(T)=H_{c2}^{(u)}(T=0) (1-T/T_{cu})^{\alpha (u)},
\label{eq05}
\end{equation}
with an exponent $\alpha (u)$ being larger than 1.  The least-squares fit
of the data of $H_{c2}^{(u)}$ vs $T$ to Eq.(\ref{eq05}) yields the
parameters $H_{c2}^{(u)}(0) = 17.1 \pm 1.0$ kOe, $T_{cu} = 9.1 \pm 0.1$ K,
and $\alpha (u) = 1.35 \pm 0.06$.  The values of $T_{cu}$ and
$H_{c2}^{(u)}(0)$ for Co$_{0.33}$Ta$_{2}$S$_{2}$C are slightly larger than
those for the pristine Ta$_{2}$S$_{2}$C, respectively; $T_{cu} = 8.9 \pm
0.1$ K and $H_{c2}^{(u)}(0) = 14.0 \pm 0.5$ kOe for
Ta$_{2}$S$_{2}$C.\cite{Walter2004} In Co$_{0.33}$Ta$_{2}$S$_{2}$C, the
superconductivity occuring in the Ta-C layer may compete with the possible
charge density wave (CDW) effect occuring in the 1T-TaS$_{2}$ layered
structure.\cite{DiSalvo1980} The slight enhancement of $T_{cu}$ in
Co$_{0.33}$Ta$_{2}$S$_{2}$C may be related to a suppression of the possible
CDW.

The coherence length $\xi$ and the magnetic penetration depth $\lambda$ are
related to $H_{c1}(0)$ and $H_{c2}(0)$ through relations $H_{c1}(0) =
(\Phi_{0}/4\pi\lambda^{2}) \ln(\lambda/\xi)$ and $H_{c2}(0) =
\Phi_{0}/(2\pi\xi^{2})$,\cite{Ketterson1999} where $\Phi_{0}$ ($= 2.0678
\times 10^{-7}$ Gauss cm$^{2}$) is the fluxoid.  When the values of
$H_{c1}(0)$ (= 68 Oe) and $H_{c2}^{(u)}(0)$ (= 17.1 kOe) are used, the
values of the Ginzburg-Landau parameter $\kappa$ ($= \lambda /\xi)$,
$\lambda$ and $\xi$ can be estimated as $\kappa = 19.3 \pm 0.3$, $\xi = 80
\pm 10 \AA$ and $\lambda = 1560 \pm 100 \AA$ for
Co$_{0.33}$Ta$_{2}$S$_{2}$C.

\subsection{\label{resultE}Ni$_{0.25}$Ta$_{2}$S$_{2}$C and
Cu$_{0.60}$Ta$_{2}$S$_{2}$C}

\begin{figure}
\includegraphics[width=7.0cm]{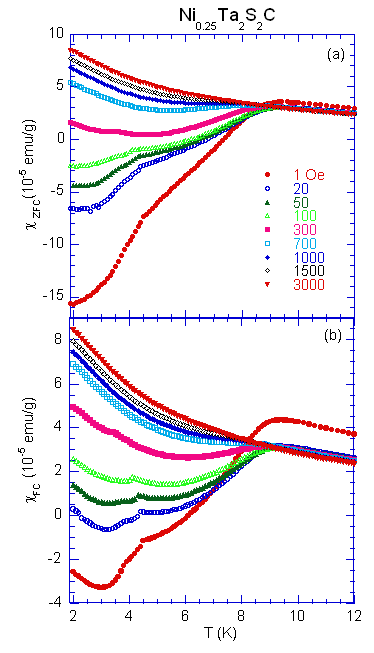}%
\caption{\label{fig14}(Color online) $T$ dependence of (a) $\chi_{ZFC}$ and
(b) $\chi_{FC}$ for Ni$_{0.25}$Ta$_{2}$S$_{2}$C at various $H$.}
\end{figure}

\begin{figure}
\includegraphics[width=7.0cm]{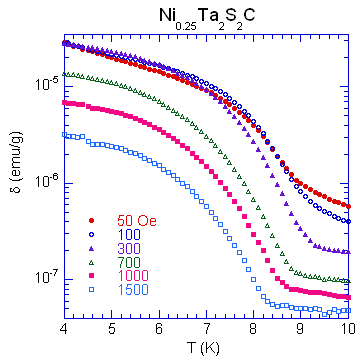}%
\caption{\label{fig15}(Color online) $T$ dependence of $\delta$ for
Ni$_{0.25}$Ta$_{2}$S$_{2}$C at various $H$.  Note that it is difficult to
derive the value of $T$ at which $\delta$ tends to zero for each $H$, from
this figure since the value of $\delta$ is plotted in a logarithmic scale,
as a function of $T$.}
\end{figure}

\begin{figure}
\includegraphics[width=8.5cm]{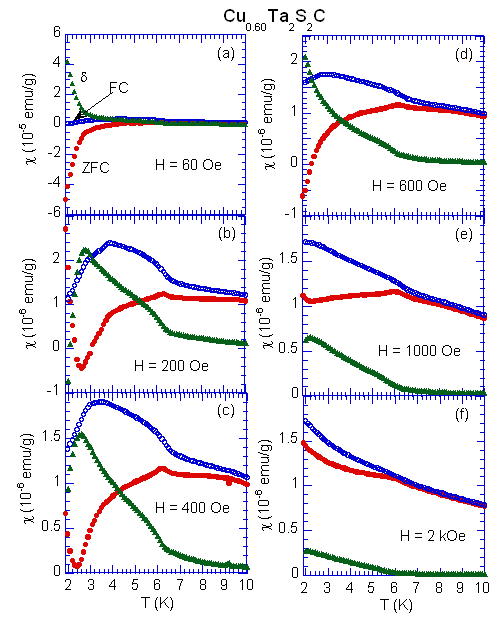}%
\caption{\label{fig16}(Color online) (a) - (f) $T$ dependence of
$\chi_{ZFC}$, $\chi_{FC}$, and $\delta$ at various $H$ for
Cu$_{0.60}$Ta$_{2}$S$_{2}$C.}
\end{figure}

\begin{figure}
\includegraphics[width=7.0cm]{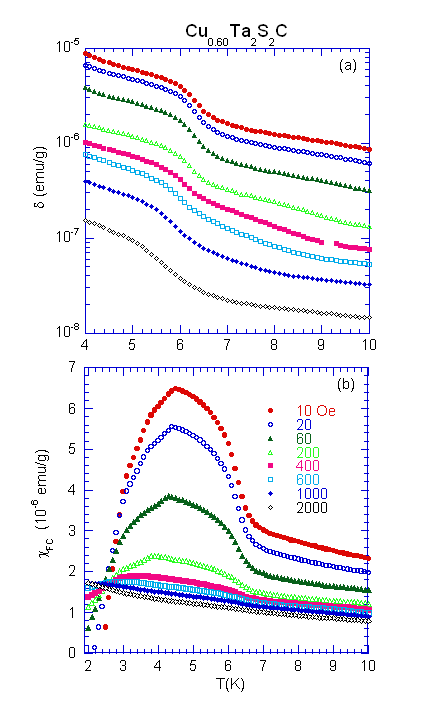}%
\caption{\label{fig17}(Color online) $T$ dependence of (a) $\delta$ and (b)
$\chi_{FC}$ at various $H$ for Cu$_{0.60}$Ta$_{2}$S$_{2}$C.}
\end{figure}

In Figs.~\ref{fig14}-\ref{fig17} we show the $T$ dependence of
$\chi_{ZFC}$, $\chi_{FC}$, and $\delta$ for Ni$_{0.25}$Ta$_{2}$S$_{2}$C and
Cu$_{0.60}$Ta$_{2}$S$_{2}$C. These results are similar to those for
Co$_{0.33}$Ta$_{2}$S$_{2}$C. The negative sign of $\chi_{ZFC}$ and
$\chi_{FC}$ indicates that the superconductivity occurs in these systems. 
For Ni$_{0.25}$Ta$_{2}$S$_{2}$C, with increasing $T$ the difference
$\delta$ sharply decreases and reduces to a weakly $H$-dependent positive
value above the critical temperature $T_{cu}$($H$).  Then $T_{cu}(H)$ can
be determined directly from the data of $\delta$ vs $T$.  We note that
$T_{cu}$ thus obtained is rather different from the local minimum
temperature of d$\delta$/d$T$ as well as the local maximum temperatures of
d$\chi_{ZFC}$/d$T$ vs $T$ and d$\chi_{FC}$/d$T$ vs $T$, partly because of
the magnetic contribution from Ni$^{2+}$ spins.  Figure \ref{fig13}(b)
shows the $H$-$T$ diagram for Ni$_{0.25}$Ta$_{2}$S$_{2}$C, where
$T_{cu}(H)$ thus obtained is plotted as a function of $H$.  The
least-squares fit of the critical line to Eq.(\ref{eq05}) yields the
parameters, $H_{c2}^{(u)}(0) = 52.0 \pm 11.0$ kOe, $T_{cu} = 8.7 \pm 0.1$
K, and $\alpha (u) = 1.57 \pm 0.12$.  For Cu$_{0.60}$Ta$_{2}$S$_{2}$C, in
contrast, the difference $\delta$ gradually decreases with increasing $T$
around $ T_{cu}$, which may be due to a broad distribution of $T_{cu}$. 
The critical temperature $T_{cu}$($H$) is defined as a temperature at which
the data of d$\delta$/d$T$ vs $T$ has a local minimum.  In
Fig.~\ref{fig13}(c) we show the $H$-$T$ diagram for
Cu$_{0.60}$Ta$_{2}$S$_{2}$C, where $T_{cu}(H)$ is plotted as a function of
$H$.  For comparison, we make a plot of the temperature at which the data
of d$\chi_{ZFC}$/d$T$ vs $T$ has a local maximum, as a function of $H$. 
The least-squares fit of the critical line to Eq.(\ref{eq05}) yields the
parameters, $H_{c2}^{(u)}(0) = 15.7 \pm 2.2$ kOe, $T_{cu} = 6.4 \pm 0.1$ K,
and $\alpha (u) = 1.15 \pm 0.07$.  The value of $T_{cu}$ is lower than that
for Co$_{0.33}$Ta$_{2}$S$_{2}$C and Ni$_{0.25}$Ta$_{2}$S$_{2}$C.

\begin{table}
\caption{\label{table3}Superconducting parameters of
$M_{c}$Ta$_{2}$S$_{2}$C and host Ta$_{2}$S$_{2}$C. There are two
transitions at $T_{cl}$ and $T_{cu}$ for $M_{c}$Ta$_{2}$S$_{2}$C and host
Ta$_{2}$S$_{2}$C.}
\begin{ruledtabular}
\begin{tabular}{llll}
$M_{c}$Ta$_{2}$S$_{2}$ & $T_{c}$ (K) & $H_{c2}(0)$ (kOe) & $\alpha$\\
\hline
Fe$^{(u)}$ & 8.8 $\pm$ 0.1 & 8.3 $\pm$ 2.2 & 1.28 $\pm$ 0.14\\
Co$^{(u)}$ & 9.1 $\pm$ 0.1 & 17.1 $\pm$ 1.0 & 1.35 $\pm$ 0.06\\
Ni$^{(u)}$ & 8.7 $\pm$ 0.1 & 52.0 $\pm$ 11.0 & 1.57 $\pm$ 0.12\\
Cu$^{(u)}$ & 6.4 $\pm$ 0.1 & 15.7 $\pm$ 2.2 & 1.15 $\pm$ 0.07\\
Fe$^{(l)}$, Ni$^{(l)}$, Cu$^{(l)}$ & 4.6 $\pm$ 0.1 & 2.2 $\pm$ 0.3 & 1.25 $\pm$ 0.16\\
\hline
Ta$_{2}$S$_{2}$C$^{(l)}$ & 3.61 $\pm$ 0.01 & 7.7 $\pm$ 0.2 & \\
Ta$_{2}$S$_{2}$C$^{(u)}$ & 8.9 $\pm$ 0.1 & 14.0 $\pm$ 0.5 & 1.23 $\pm$ 0.07\\
\end{tabular}
\end{ruledtabular}
\end{table}

The critical temperatures $T_{cu}$ in $M_{c}$Ta$_{2}$S$_{2}$C are listed in
Table \ref{table3}.  If $T_{cu}$ at $c = 0$ (Ta$_{2}$S$_{2}$C) corresponds
to $T_{cu}$ ($= 8.9 \pm 0.1$ K), then the lowering of $T_{cu}$ is clearly
observed for Ni$_{0.25}$Ta$_{2}$S$_{2}$C and Cu$_{0.60}$Ta$_{2}$S$_{2}$C.
Similar behavior has been reported by Sakamaki et al.\cite{Sakamaki2001} in
Cu$_{c}$Nb$_{2}$S$_{2}$C: $T_{c} = 7.6$ K at $c = 0$ and $T_{c} = 4.8$ K
for $c = 0.70$.  In $M_{c}X_{2}$S$_{2}$C ($X$ = Nb, Ta), there is an
electron charge transfer from the $M_{c}$ atoms to the host lattice
$X_{2}$S$_{2}$C. The increase of conduction electrons leads to the lowering
of $T_{c}$.

In Fig.~\ref{fig17}, $\chi_{FC}$ shows a broad peak (or a cusp) near the
temperature $T_{cl}$ (= 4.6 K) for Cu$_{0.60}$Ta$_{2}$S$_{2}$C. This peak
shifts to the low-$T$ side with increasing $H$.  Similar behaviors are
observed in the form of either a cusp or a kink at $T_{cl}(H)$ in both
$\chi_{FC}$ and $\chi_{ZFC}$ for Ni$_{0.25}$Ta$_{2}$S$_{2}$C and
Fe$_{0.33}$Ta$_{2}$S$_{2}$C. These data of $T_{cl}(H)$ vs $H$ for
Ni$_{0.25}$Ta$_{2}$S$_{2}$C and Cu$_{0.60}$Ta$_{2}$S$_{2}$C are shown in
Fig.~\ref{fig09}(a).  The critical line $H_{c2}^{(l)}(T)$ is described by a
power law form similar to Eq.(\ref{eq05}) with $T_{cl} = 4.6 \pm 0.1$ K,
$H_{c2}^{(l)}(0) = 2.2 \pm 0.3$ kOe, and $\alpha (l) = 1.25 \pm 0.16$. 
We find that all the data of $T_{cl}$ vs $H$ fall on the critical line
$H_{c2}^{(l)}(T)$ common to our all systems.

The effective magnetic moment $P_{eff}$ of Ni$_{0.25}$Ta$_{2}$S$_{2}$C and
Cu$_{0.60}$Ta$_{2}$S$_{2}$C is much smaller than the spin-only values.  In
fact, the $g$-factors of Ni$_{0.25}$Ta$_{2}$S$_{2}$C ($g =1.64 \pm 0.02$)
and Cu$_{0.60}$Ta$_{2}$S$_{2}$C ($g = 0.324 \pm 0.003$) are much smaller
than that observed in Ni$^{2+}$ spin in stage-2 NiCl$_{2}$ graphite
intercalation compound (GIC) ($g_{a} = 2.156 \pm 0.002$ and $g_{c} = 2.096
\pm 0.002$),\cite{Suzuki1984} and Cu$^{2+}$ spin in stage-2 CuCl$_{2}$ GIC
($g_{a} = 2.08 \pm 0.01$ and $g_{c} = 2.30 \pm 0.01$).\cite{Suzuki1994}
These results suggest that the $M$ (= Ni and Cu) 3d moments are partially
delocalized, metallic type of bonding: $M$ 3d electrons are more itinerant. 
The $M$ 3d states hybridize strongly with the Ta 5d$_{z}^{2}$ which from
the conduction bands of the host compound.

\section{\label{dis}DISCUSSION}
\subsection{\label{disA}$XY$ symmetry of Fe$^{2+}$ spins}
Here we discuss the symmetry of spin Hamiltonian for
Fe$_{0.33}$Ta$_{2}$S$_{2}$C with $T_{N}=8.7$ K, where the intraplanar
exchange interaction is antiferromagnetic.  For simplicity, we assume that
the parameters of Fe$^{2+}$ spins for Fe$_{0.33}$Ta$_{2}$S$_{2}$C is the
same as that of FeCl$_{2}$; $x = \delta_{0}/\lambda^{\prime} = -1.27$, $p =
1.67$, $q = 1.41$, $(p/q)^{2} - 1 = 0.40$, $g_{c}^{(0)} = 3.972$,
$g_{a}^{(0)} = 3.195$, $k = 0.95$, $\lambda^{\prime} = -67$ cm$^{-1}$, and
$D = 7.0$ cm$^{-1} = 10$ K (See Sec.~\ref{back}).  In the present work, we
have measured the average value of $g$ for the powdered sample, which is
described by either $g_{av1} = (g_{c} + 2g_{a})/3$ or $g_{av2} =
[(g_{c}^{2} + 2g_{a}^{2})/3]^{1/2}$, depending on the form of
$g$:\cite{Abragam1970} $g_{av1} = \langle \cos^{2}(\theta)\rangle g_{c} +
\langle\sin^{2}(\theta)\rangle g_{a}$ or $g_{av2}^{2} =
\langle\cos^{2}(\theta)\rangle g_{c}^{2} + \langle\sin^{2}(\theta)\rangle
g_{a}^{2}$, where $\theta$ is the angle between the magnetic field
direction and the $z$ axis, and $\langle\cos^{2}(\theta)\rangle = 1/3$ and
$\langle\sin^{2}(\theta)\rangle = 2/3$.  If $g_{a} = g_{a}^{(0)} + \Delta g
= 2.91$ and $g_{c} = g_{c}^{(0)} = 3.68$ where $\Delta g = -0.29$ is
appropriately chosen, then the $g$-value can be estimated as either
$g_{av1} = 3.17$ or $g_{av2} = 3.19$, which are the same as our
experimental value $g = 3.16 \pm 0.02$ for Fe$_{0.33}$Ta$_{2}$S$_{2}$C. The
spin anisotropy parameter $D_{eff}$ can be estimated as $D_{eff} = D (s
-1/2) + (3\Theta /(s+1)) (p^{2}-q^{2})/q^{2}) \approx 5 +(-29.5/2) 0.4 =
-0.9$ K with $s = 1$.  The negative sign of $D_{eff}$ suggests the
easy-plane type ($XY$) spin anisotropy of the system: the spins tend to lie
in the $c$ plane (perpendicular to the $c$ axis).  Thus the Fe$^{2+}$ spins
magnetically behave like an $XY$-like AF on the triangular lattice.

Here it is interesting to point out that the magnetic properties of
Fe$_{0.33}$Ta$_{2}$S$_{2}$C are similar to those of 2$H$-Fe$_{c}$NbSe$_{2}$
which belongs to the same universality class: $XY$-like AF on the
triangular lattice.  The magnetic properties of 2$H$-Fe$_{c}$NbSe$_{2}$
have been extensively studied by Hillenius and Coleman.\cite{Hillenius1979}
Their results are summarized as follows.  The Curie-Weiss temperature
$\Theta$ of 2$H$-Fe$_{c}$NbSe$_{2}$ is negative for $0.05 \leq c \leq
0.33$.
At $c = 0.12$ the susceptibility $\chi_{a}$ shows a sharp peak at 12 K with
the maximum of 0.092 emu/Fe mole.  The susceptibility $\chi_{a}$ is larger
than $\chi_{c}$, suggesting $XY$ anisotropy: $g_{a}= 2.40$ and $g_{c} =
2.62$.  At $c = 0.33$ the susceptibility $\chi_{a}$ shows a sharp peak at
135 K, suggesting the existence of AF long range order.  The maximum
susceptibility is 0.02 emu/Fe mole.

In Fe$_{0.33}$Ta$_{2}$S$_{2}$C the interplanar interaction $J^{\prime}$ is
considered to be ferromagnetic but is extremely weak compared with the AF
intraplanar exchange interaction.  For the 2D $XY$ spin system, a stable
spin structure is realized as a 120$^\circ$ spin structure
($\sqrt{3}\times\sqrt{3}$)R30$^\circ$ commensurate spin structure), where
spins rotate successively by 120$^\circ$.  Sakakibara\cite{Sakakibara1984} has pointed out
that the 120$^\circ$ spin structure becomes unstable by an infinitesimal
interplanar exchange interaction $J^{\prime}$, and that the incommensurate
magnetic structures appear.

\subsection{\label{disB}Ising symmetry of Co$^{2+}$ spins}
Next we discuss the spin symmetry of Co$^{2+}$ spin in
Co$_{0.33}$Ta$_{2}$S$_{2}$C, where the intraplanar exchange interaction is
antiferromagnetic.  For simplicity, we assume that the parameters of
Co$^{2+}$ spins for Co$_{0.33}$Ta$_{2}$S$_{2}$C is the same as that of
CoCl$_{2}$: $x = \delta_{0}/\lambda^{\prime} = 1.68$, $p = 1.36$, $q
=1.81$, ($p/q)^{2} - 1 = -0.44$, $g_{c}^{(0)} = 3.23$, $g_{a}^{(0)} =
4.69$, and $k = 0.90$.  If $g_{a} = g_{a}^{(0)} + \Delta g = 3.15$ and
$g_{c} = g_{c}^{(0)} + \Delta g = 4.61$ where $\Delta g = -0.08$ is
appropriately chosen, then the $g$-value can be estimated as either
$g_{av1} = 4.12$ or $g_{av2} = 4.18$, which are the same as our
experimental value $g = 4.12$ $\pm 0.05$ for Co$_{0.33}$Ta$_{2}$S$_{2}$C.
For Co$_{0.33}$Ta$_{2}$S$_{2}$C, the spin anisotropy parameter can be
estimated as $2zsJ_{A} = (3\Theta/(s+1)) (p^{2}-q^{2})/q^{2}) \approx 23.1$
K with $s = 1/2$ and $\Theta = -26.3$ K. The positive sign of $J_{A}$
suggests the Ising spin anisotropy of the system: the spins tend to align
along the $c$ axis.  The AF phase transition occurs at $T_{N}$ (= 18.6 K). 
Note that $T_{N}$ is lower than $\vert\Theta\vert$ (= 26.3 K), because of
the deviation of molecular field theory.  The ratio $\eta$
($=T_{N}/\vert\Theta\vert$) is equal to 0.70, which is close to $\eta$ (=
0.65) for CoCl$_{2}$.\cite{Starr1940,Hutchings1973} The Co$^{2+}$ spins
magnetically behave like an Ising-like AF on the triangular lattice.  A
spin frustration effect arising from a macroscopic number of degenerate
ground states suppresses the growth of the long range spin ordering. 
CsCoCl$_{3}$,\cite{Mekata1977} which is one of the best known systems,
undergoes two phase transitions at $T_{N1}$ (= 21 K) and $T_{N2}$ (= 9 K). 
In the intermediate phase between $T_{N1}$ and $T_{N2}$, spins on two of
the three sublattices order antiferromagnetically and the rest remains
paramagnetic, forming a so-called partially disordered (PD) phase.  Two
peaks of both $\chi_{ZFC}$ and $\chi_{FC}$ in Co$_{0.33}$Ta$_{2}$S$_{2}$C
at $T_{N}$ and $T_{cu}$ may corresponds to $T_{N1}$ and $T_{N2}$, although
$T_{cu}$ is regarded as a superconducting transition point (see
Sec.~\ref{resultD}).  We note that similar behavior is also observed in
(CeS)$_{1.16}$[Fe$_{0.33}$(NbS$_{2}$)$_{2}$],\cite{Michioka2002} which is
also a quasi 2D Ising AF on the triangular lattice.  This system undergoes
phase transitions at $T_{N1}$ (= 22 K) and $T_{N2}$ (= 15 K).  The
intermediate phase between $T_{N1}$ and $T_{N2}$ is a PD phase.

\subsection{\label{disC}Nature of superconductivity}
What is the nature of the superconducting transition at $T_{cl}$ (= 4.6 K)? 
Both $\chi_{ZFC}$ and $\chi^{\prime}$ show an anomaly at $T_{cl}$, which is
independent of the kind of $M$ in $M_{c}$Ta$_{2}$S$_{2}$C. The critical
line $T_{cl}(H)$ is almost the same for $M$ = Fe, Ni, and Cu.  The value of
$T_{cl}$ is slightly larger than that of $T_{cl}$ for the host
Ta$_{2}$S$_{2}$C, partly because of a suppression of the possible CDW. This
superconductivity at $T_{cl}$ can coexist with the AF long range spin order
below $T_{N}$ in Fe$_{0.33}$Ta$_{2}$S$_{2}$C and
Co$_{0.33}$Ta$_{2}$S$_{2}$C. As listed in Table \ref{table3}, the exponent
$\alpha$ for Co$_{0.33}$Ta$_{2}$S$_{2}$C and Ni$_{0.33}$Ta$_{2}$S$_{2}$C is
on the same order as those for quasi 2D superconductors such as Pd-MG
(metal graphite) ($\alpha = 1.43 \pm 0.05$ and $T_{c} = 3.82 \pm 0.04$
K)\cite{Suzuki2004a} and Sn-MG ($\alpha = 1.43 \pm 0.05$ and $T_{c} = 3.90
\pm 0.08$ K).\cite{Suzuki2004b} In these MG's, metal (Pd and Sn) layers,
which are sandwiched between adjacent graphene sheets, exhibit
superconductivity.  Note that the exponent $\alpha$ of
$M_{c}$Ta$_{2}$S$_{2}$C is close to that (= 1.50) predicted from the de
Almeida and Thouless\cite{AT1978} theory for spin glass behaviors. 
This result suggests that the critical line may correspond to an
irreversibility line where the difference $\delta$ starts to increase with
decreasing $T$.

\section{CONCLUSION}
$M_{c}$Ta$_{2}$S$_{2}$C exhibits a variety of superconducting and magnetic
properties. 
Fe$_{0.33}$Ta$_{2}$S$_{2}$C, which is a quasi 2D $XY$-like antiferromagnet
on the triangular lattice, undergoes an AF transition at $T_{N}$ (= 117 K). 
The irreversible effect of magnetization occurs below $T_{N}$, reflecting
the 2D frustrated nature of the system.  The AF phase coexists with the
superconducting phase below $T_{cu}$ (= 8.8 K). 
Co$_{0.33}$Ta$_{2}$S$_{2}$C, which is a quasi 2D Ising antiferromagnet on
the triangular lattice, shows an AF transition at $T_{N} = 18.6$ K. The AF
phase coexists with the superconducting phase below $T_{cu}$ = 9.1 K. Both
Ni$_{0.25}$Ta$_{2}$S$_{2}$C and Cu$_{0.60}$Ta$_{2}$S$_{2}$C are
superconductor with $T_{cu}$ = 8.7 K and 6.4 K, respectively. 
$M_{c}$Ta$_{2}$S$_{2}$C also exhibits another superconducting transition at
$T_{cl}$ (= 4.6 K), which is common to our all systems and is a little
higher than that of the host Ta$_{2}$S$_{2}$C ($T_{cl}$ = 3.61 K).

\begin{acknowledgments}
The authors are grateful to Pablo Wally, Technical University of Vienna,
Austria (now Littlefuse, Yokohama, Japan) for providing us with the
samples.  One of the authors (J.W.) acknowledges financial support from the
Ministry of Cultural Affairs, Education and Sport, Japan, under the grant
for young scientists no.  70314375 and from Kansai Invention Center, Kyoto,
Japan.
\end{acknowledgments}

\end{document}